\documentclass[twocolumn,english,aps,prd,showpacs]{revtex4-1}
\usepackage[T1]{fontenc}
\usepackage[latin9]{inputenc}
\usepackage{xcolor}
\usepackage{pdfcolmk}
\usepackage{amsmath}
\usepackage{amssymb}
\usepackage{graphicx}
\PassOptionsToPackage{normalem}{ulem}
\usepackage{ulem}

\makeatletter

\providecolor{lyxadded}{rgb}{0,0,1}
\providecolor{lyxdeleted}{rgb}{1,0,0}

\usepackage{babel}

\makeatother

\usepackage{babel}
\begin{document}

\title{Renormalization Group Approach to Stability of Two-dimensional Interacting
Type-II Dirac Fermions}

\author{Ze-Min Huang}

\address{Department of Physics, The University of Hong Kong, Pokfulam Road,
Hong Kong, China}

\author{Jianhui Zhou}

\address{Department of Physics, The University of Hong Kong, Pokfulam Road,
Hong Kong, China}

\author{Shun-Qing Shen}
\email{sshen@hku.hk}

\address{Department of Physics, The University of Hong Kong, Pokfulam Road,
Hong Kong, China}

\date{\today }
\begin{abstract}
The type-II Weyl/Dirac fermions are a generalization of conventional
or type-I Weyl/Dirac fermions, whose conic spectrum is tilted such
that the Fermi surface becomes lines in two dimensions, and surface
in three dimensions rather than discrete points of the conventional
Weyl/Dirac fermions. The mass-independent renormalization group calculations
show that the tilting parameter decreases monotonically with respect
to the length scale, which leads to a transition from two dimensional
type-II Weyl/Dirac fermions to the type-I ones. Because of the non-trivial
Fermi surface, a photon gains a finite mass partially via the chiral
anomaly, leading to the strong screening effect of the Weyl/Dirac
fermions. Consequently, anisotropic type-II Dirac semimetals become
stable against the Coulomb interaction. This work provides deep insight
into the interplay between the geometry of Fermi surface and the Coulomb
interaction.
\end{abstract}
\maketitle

\section{Introduction}

Three-dimensional (3D) Weyl semimetals are topological states of quantum
matter and can be regarded as 3D analogues of graphene \cite{Wan2011prb,WengHM2016jpcm}.
Their conduction and valence bands with linear dispersion touch each
other at a finite number of points, called the Weyl nodes, in the
3D Brillouin zone. These Weyl nodes can be viewed as magnetic monopoles
in momentum space \cite{Xiao2010RMP,Volovikbook}, which lead to various
novel electromagnetic responses such as the chiral anomaly \cite{Nielsen1983plb,Aji2012PRB,WangZ2013prb,LiuCX2013PRB,Goswami2013PRB,Parameswaran2014prx},
chiral magnetic effect \cite{Fukushima2008CME,Grushin2012PRD,Zyuzin2012PRB,Zhou2013CPL,Vazifeh2013prl,Chang2015PRB1,MaPesin2015cme,zhong2016PRL},
and exotic magnetoresistance \cite{SonSpivak13prb,GorbarPRB2014,Burkov2014prl,LuHZ2015prb}.
Very recently, a new kind of Weyl/Dirac semimetals with the tilted
conic spectrum, named type-II Weyl/Dirac semimetals, have been predicted
in a series of materials \cite{Soluyanov2015nature,Sun2015PRB,KoepernikTaIrTe,autes2016PRL},
in which the Fermi surface crossing the Weyl nodes is lines in two
dimensions \cite{Goerbig2008PRB} and surface in three dimensions
\cite{Soluyanov2015nature} with a finite density of states (DOS).
Meanwhile, many experiments have made great efforts to characterize
type-II Weyl/Dirac semimetals by angle-resolved photoemission spectroscopy
\cite{Huang2016NM,xu2016LaAlGe,WangCL2016PRB}. This nontrivial Fermi
surface could lead to an exotic magnetic-optical response \cite{Yuzm2016PRL,Tchoumakov2016PRL,Udagawa2016PRL}
and unconventional magnetic breakdown \cite{Obrien2016prl}.

Coulomb interaction plays a crucial role in understanding the properties
and stability of the Fermi surface in both two-dimensional (2D) \cite{Kotov2012RMP}
and 3D Dirac/Weyl semimetals \cite{Abrikosov1971,Zhou2015PRB,Hofmann2015PRB}.
For type-I 2D Dirac semimetals, the vanishing DOS at Fermi points
does not screen the Coulomb interaction sufficiently \cite{Kotov2012RMP}.
The renormalization group (RG) \cite{Shankar1994RMP,Polchinski1992arxiv}
calculations show that the Coulomb interaction renormalizes the effective
velocity and makes it diverge logarithmically \cite{Gonzalez1999PRB,Sondt2007prb}.
One solution to this unphysical divergence is to take into account
the relativistic effect in the full quantum electrodynamics level
\cite{Isobe2016PRL} such that the velocity of fermions is renormalized
up to the speed of light \cite{Isobe2012jpsj}. In this work, we study
2D type-II Dirac semimetals that possess an extended Fermi surface
(two lines) and finite DOS even when the Fermi surface crosses the
Weyl nodes. It is natural to wonder how this nontrivial geometry of
Fermi lines interplays with both the Coulomb interaction and the finite
DOS.

It has been shown that under the unscreened Coulomb interaction, the
tilting parameter is a monotonic decreasing function of length scale
such that the 2D type-II Dirac fermions transit to the type-I ones.
Because of the non-trivial Fermi surface, a photon gains a finite
mass partially via the chiral anomaly, giving rise to a strong screening
effect. Consequently, anisotropic type-II Dirac fermions become stable
against the Coulomb interaction. In addition, the logarithmic divergence
of Fermi velocity is substantially suppressed. This work sheds new
light on the interaction effect of 2D type-II Dirac semimetals.

\begin{figure}
\includegraphics[scale=0.4]{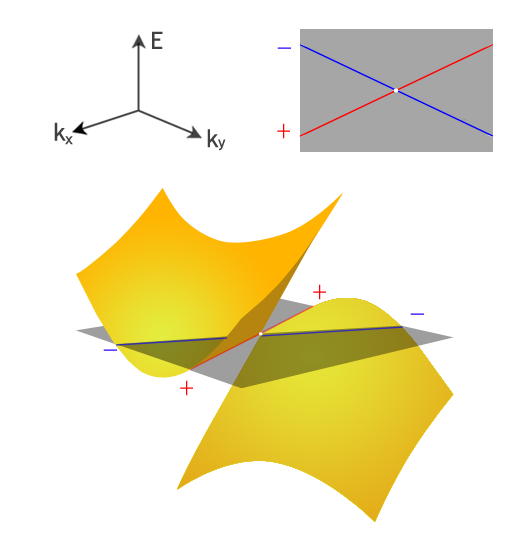}

\caption{The energy bands and Fermi surface of 2D type-II Dirac fermions. The
dark plane is the iso-energy surface crossing the Weyl node located
by the white intersection point. The red and blue lines labeled by
$\pm$ denote the Fermi lines. }
\end{figure}

\section{model for 2d tilted Dirac fermions}

In the present work, we shall focus on the 2D tilted Dirac fermions
that respect the inversion symmetry or time reversal symmetry to prevent
the band-gap opening. Actually, the quasi-2D conducting organic compound
$\alpha-\left(\mathrm{BEDT-TTF}\right)_{2}\mathrm{I}_{3}$ supports
these tilted Dirac fermions \cite{Bender1984mclc,Katayama2006jpsj,Hirata2016NC}.
We start with the following Hamiltonian for 2D type-II Dirac fermions,
in which the conic spectrum is tilted along the $x$-axis (we set
$\hbar=1$) \cite{Goerbig2008PRB},

\begin{equation}
H_{0}=v_{x}p_{x}\sigma^{1}+v_{y}p_{y}\sigma^{2}+wv_{x}p_{x},\label{eq1}
\end{equation}
where $w$ refers to a tilting parameter for type-II Dirac fermions
with $|w|>1$ and for type-I ones otherwise. $v_{x}$ and $v_{y}$
denote for the velocity along the $x$-axis and $y$-axis, respectively.
For simplicity, we denote the ratio of two velocities by $\eta=v_{x}/v_{y}$.
$\sigma^{i}$ with $i=1,2$ are the Pauli matrices. The energy dispersions
of Eq. (\ref{eq1}) are of the form

\begin{equation}
E_{\pm}=wv_{x}p_{x}\pm\sqrt{v_{x}^{2}p_{x}^{2}+v_{y}^{2}p_{y}^{2}}.\label{Edisp}
\end{equation}
There are two Fermi lines at $E_{\pm}=0$ in the $p_{x}-p_{y}$ plane,
described by $p_{y}=\pm\tilde{w}\eta p_{x}$ with $\tilde{w}=\sqrt{w^{2}-1}$.
$\pm$ refers to the sign of slope of each Fermi line as shown in
Fig. 1. Note that the energy dispersion in Eq. $\left(\ref{Edisp}\right)$
is only valid for a finite region near each Weyl node, we thus introduce
a momentum cut-off $\Lambda$ along the $p_{x}$-axis. Since the Fermi
surface is not point-like, one needs to expand the Hamiltonian in
Eq. (\ref{eq1}) around the Fermi lines to capture the low-energy
properties \cite{Polchinski1992arxiv,Shankar1994RMP}. Namely, we
decompose the momentum into two parts: the momentum parallel to the
Fermi lines $k_{F}$ and the momentum perpendicular to the Fermi lines
$\tilde{p}$,
\begin{eqnarray}
k_{F}^{\pm} & = & (p_{\parallel},\pm\tilde{w}\eta p_{\parallel}),\nonumber \\
\tilde{p}^{\pm} & = & (\pm\tilde{w}\eta p_{\perp},-p_{\perp})\label{eqMomentum-1}
\end{eqnarray}
with $J=1+\tilde{w}^{2}\eta^{2}$ being the corresponding Jacobian.
$p_{\Vert}$ and $p_{\bot}$ are the $x$-component of $k_{F}^{\pm}$
and $y$-component of $\tilde{p}^{\pm}$, respectively, where the
subscript $\Vert$ $\left(\bot\right)$ denotes for momentum parallel
(perpendicular) to the Fermi lines. In terms of $k_{F}^{\pm}$ and
$\tilde{p}^{\pm}$, the Hamiltonian in Eq. (\ref{eq1}) can be recast
as
\begin{equation}
H_{0}=H_{k_{F}^{\pm}}+H_{\tilde{p}^{\pm}},
\end{equation}
where $H_{k_{F}^{\pm}}$ and $H_{\tilde{p}^{\pm}}$ only depend on
$p_{\parallel}$ and $p_{\ensuremath{\perp}}$, respectively. The
corresponding Lagrangian for electrons can be directly obtained from
$H_{k_{F}^{\pm}}$ and $H_{\tilde{p}^{\pm}}$,

\begin{align}
\mathcal{L}_{e} & =\Sigma_{s=\pm}\Psi_{s}^{\dagger}\left[p_{0}-\left(H_{k_{F}^{s}}+H_{\tilde{p}^{s}}\right)p_{\perp}\right]\Psi_{s}\nonumber \\
 & =\Psi_{+}^{\dagger}(p_{\bot})\left[p_{0}-\left(w\tilde{w}\eta^{2}+\tilde{w}\eta^{2}\sigma^{1}-\sigma^{2}\right)v_{y}p_{\perp}\right]\Psi_{+}(p_{\bot})\nonumber \\
 & +\Psi_{-}^{\dagger}\left(-p_{\bot}\right)\left(-i\sigma^{2}\right)\left[p_{0}-\left(w\tilde{w}\eta^{2}-\tilde{w}\eta^{2}\sigma^{1}+\sigma^{2}\right)v_{y}p_{\perp}\right]\nonumber \\
 & \times\left(i\sigma^{2}\right)\Psi_{-}\left(-p_{\bot}\right),
\end{align}
where $\Psi_{s}$ with $s=\pm$ denote for fermions on two different
Fermi lines and satisfy the relation $\Psi_{\pm}^{\dagger}H_{k_{F}^{\pm}}\Psi_{\pm}=0$
due to zero Fermi energy. After introducing a new spinor

\begin{equation}
\bar{\Psi}\left(p_{\bot}\right)=\left(\Psi_{-}^{\dagger}\left(-p_{\bot}\right)(-i\sigma^{2}),\Psi_{+}^{\dagger}\right)\gamma^{0},
\end{equation}
the action can be recast as
\begin{eqnarray}
S_{0} & = & \int\frac{dp_{0}dp_{\parallel}dp_{\perp}}{\left(2\pi\right)^{3}}J(\mathcal{L}_{e}+\mathcal{L}_{\gamma}+\mathcal{L}_{I}),\label{eqFeynmanRule}
\end{eqnarray}
where $\mathcal{L}_{e}$, $\mathcal{L}_{\gamma}$ and $\mathcal{L}_{I}$
denote the Lagrangians for electrons, photons and the interaction
between electrons and photons, respectively. Their specific expressions
are given as

\begin{eqnarray}
\mathcal{L}_{e} & = & \bar{\Psi}\left[p_{0}\gamma^{0}-\left(w\tilde{w}\eta^{2}\gamma^{0}+\tilde{w}\eta^{2}\gamma^{1}-\gamma^{2}\right)v_{y}p_{\perp}\right]\Psi,\nonumber \\
\mathcal{L}_{\gamma} & = & \int\frac{dp_{z}}{4\pi}\phi\left[J\left(p_{\parallel}^{2}+p_{\perp}^{2}\right)+p_{z}^{2}\right]\phi,\nonumber \\
\mathcal{L}_{I} & = & \bar{\Psi}\left(-e\gamma^{0}\phi\right)\Psi,\label{Lagepi}
\end{eqnarray}
where $\gamma^{\mu}$ with $\mu=0,1,2$ are the $4\times4$ Dirac
gamma matrices including the extra two degrees of freedom for the
two Fermi lines. $\phi$ refers to the field of photons. By use of
the Hubbard-Stratonovich transformation, one obtains both $\mathcal{L}_{\gamma}$
and $\mathcal{L}_{I}$ from the Coulomb interaction,
\[
(\bar{\Psi}\gamma^{0}\Psi)(-p)V(q)(\bar{\Psi}\gamma^{0}\Psi)(p),
\]
which describes both the inter- and the intra-Fermi line interactions
and satisfies the restriction given by the Fermi surface at tree level.
$V(q)=e^{2}/\sqrt{q_{x}^{2}+q_{y}^{2}}$ is the Fourier transformation
of a bare 2D Coulomb potential. Additionally, $\mathcal{L}_{\gamma}$
is actually the non-relativistic limit of the action $-\frac{1}{4}\int\frac{d^{4}p}{(2\pi)^{4}}F^{\mu\nu}F_{\mu\nu}$
in the Feynman-'t Hooft gauge with $F_{\mu\nu}$ being the electromagnetic
field strength tensor. $\int\frac{dp_{z}}{2\pi}$ in $\mathcal{L}_{\gamma}$
comes from \cite{Sondt2007prb}
\begin{equation}
\frac{1}{2\sqrt{p_{x}^{2}+p_{y}^{2}}}=\int\frac{dp_{z}}{2\pi}\frac{1}{p_{x}^{2}+p_{y}^{2}+p_{z}^{2}},
\end{equation}
where we have set the speed of light $c=1$ for simplicity. It should
be noted that when $\tilde{w}=0$, the Fermi surface at $E_{\pm}=0$
becomes a straight line and the Lagrangian for electrons reduces to
be
\begin{equation}
\mathcal{L}_{e}=\bar{\Psi}\left(p_{0}\gamma^{0}+\gamma^{2}v_{y}p_{\perp}\right)\Psi,
\end{equation}
which is nothing but the famous Schwinger model for the $\left(1+1\right)$D
massless Dirac fermions \cite{Schwinger1962PR}. Under this circumstance,
the photon gains a finite mass due to the chiral anomaly \cite{Adler1969pr,Bell1969IINCA,roskies1981prd}.
A finite photon mass usually leads to a strong screening effect in
the long-range Coulomb interaction. In fact, the screening effect
shall play a critical role in suppressing the logarithmic divergence
of Fermi velocity and stabilizing the 2D anisotropic type-II Dirac
fermions.

\section{The screening effect from analysis of random phase approximation }

The finite DOS and the extended Fermi surface in the type-II Weyl/Dirac
semimetals remind us of the role of the screening effect. By calculating
the vacuum polarization diagram in Fig. (2b) (the detailed derivations
are given in Appendix ), one finds

\begin{align}
\Pi^{00}\left(p_{0},\vec{p}\right) & =-i\int J\mathrm{tr}\left(\Gamma_{0}G_{e0}\Gamma_{0}G_{e0}\right)\nonumber \\
 & =\frac{4e^{2}}{\pi}\frac{\Lambda^{2}\mathcal{C}^{-1}\left(1+\tilde{w}^{2}\eta^{2}\right)p_{\perp}^{2}}{(p_{0}+w^{2}\tilde{w}^{2}\eta^{4}p_{\perp})^{2}-\mathcal{C}^{-2}p_{\perp}^{2}},\label{Pi00}
\end{align}
where $\Gamma_{0}$ and $G_{e0}$ are the vertex and electron's propagator,
respectively, and $\mathcal{C}=1/\sqrt{v_{y}^{2}\left(\tilde{w}^{2}\eta^{4}+1\right)}$.
Several remarks on $\Pi^{00}\left(p_{0},\vec{p}\right)$ are in order
here. First,
\begin{equation}
\Pi^{00}(0,\vec{p})=\frac{4e^{2}}{\pi}\frac{\Lambda^{2}\mathcal{C}^{-1}\left(1+\tilde{w}^{2}\eta^{2}\right)}{w^{4}\tilde{w}^{4}\eta^{8}-\mathcal{C}^{-2}}
\end{equation}
 is a momentum independent constant and relates to the quadratic potential
for photon mass. Second, a negative quadratic potential usually indicates
an instability of the system \cite{Srednicki2007cambridge}. In addition,
in the limit of $\tilde{w}\rightarrow0$, one finds
\begin{equation}
\lim_{\tilde{w}\rightarrow0}\frac{\Pi^{00}(p_{0},\vec{p})}{2\Lambda^{2}\left(\tilde{w}^{2}\eta^{2}+1\right)\mathcal{C}}=\frac{2e^{2}}{\pi v_{y}}\frac{v_{y}^{2}p_{\bot}^{2}}{p_{0}^{2}-v_{y}^{2}p_{\bot}^{2}},
\end{equation}
which is the same as the one of the Schwinger model that is effectively
induced by (1+1)D chiral anomaly \cite{Schwinger1962PR,roskies1981prd}.
In the remainder of this paper, we shall focus on the case with negative
$\Pi^{00}(0,\vec{p})$ \footnote{Under certain conditions, the static density-density response function
of type-II Dirac semimetals becomes positive $\chi_{nn}>0$. The corresponding
electronic compressibility becomes negative and can be detected experimentally
\cite{Eisenstein1992PRL}}.

The extended Fermi surface makes the random phase approximation (RPA)
reasonable \cite{Shankar1994RMP}. Following the standard procedure,
one finds the effective Coulomb potential within the RPA:
\begin{equation}
V_{\mathrm{eff}}(\vec{p})=V\left(\vec{p}\right)/\epsilon\left(\vec{p}\right),
\end{equation}
where the static dielectric function $\epsilon\left(\vec{p}\right)$
is given as
\begin{equation}
\epsilon(\vec{p})=1-V(\vec{p})\text{\ensuremath{\Pi}}^{00}(0,\vec{p}).
\end{equation}
According to $\epsilon(\vec{p})$, for the long-range interaction
with small momentum $p$, the magnitude of dielectric function is
much larger than 1, which reduces the strength of the Coulomb interaction.
On the other hand, for the short-range interaction with large momentum,
the magnitude of the dielectric function is almost 1, so the Coulomb
interaction is almost not modified. That is, the long-range interaction
is screened due to the extended Fermi surface, however the short-range
one is not screened. Since the renormalized effect upon the one-body
operator induced by the short-range interaction can be absorbed by
the chemical potential \cite{Shankar1994RMP}, the analysis above
implies both the Fermi velocities $v_{x}$ and $v_{y}$, and the tilting
parameter $w$ are intact under the screened Coulomb interaction.

\begin{figure}
\includegraphics[scale=0.4]{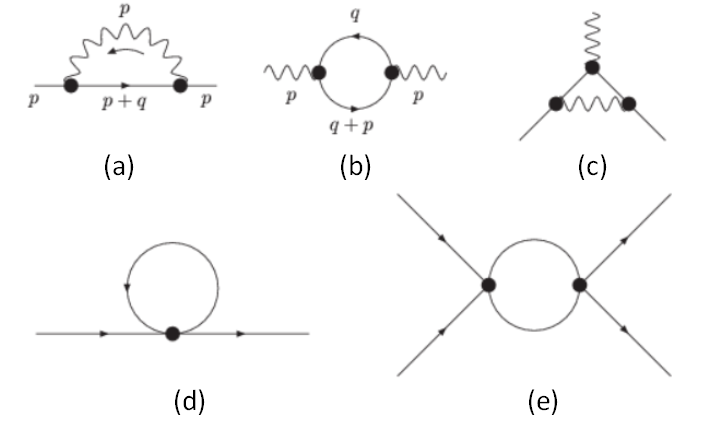}\caption{Relevant Feynman diagrams: (a), (b) and (c) are the diagrams without
decoupling effect, whereas (d) and (e) are the corresponding diagrams
with decoupling effect. }
\end{figure}

\section{renormalization group analysis}

Now we turn to consider the running effect of coupling constants under
the influence of the Coulomb interaction. We employ the dimensional
regularization and the modified minimal subtraction ($\overline{MS}$)
scheme \cite{Srednicki2007cambridge} to derive the corresponding
RG equations, with the help of Feynman's rules obtained from the Lagrangian
in Eq. (\ref{Lagepi}). The bare action is given as

\begin{eqnarray}
\mathcal{L}_{\text{bare}} & = & \bar{\Psi}[(Z_{\Psi}p_{0}-Z_{w}w\tilde{w}\eta^{2}v_{y}p_{\perp}-Z_{e}e\phi)\gamma^{0}\nonumber \\
 &  & -Z_{\tilde{w}\eta^{2}v_{y}}\tilde{w}\eta^{2}v_{y}p_{\perp}\gamma^{1}+Z_{v_{y}}v_{y}p_{\perp}\gamma^{2}]\Psi,
\end{eqnarray}
where $Z$'s are the renormalization coefficients. Because of the
U(1) symmetry, the Ward identity is valid, which implies the relation:
$Z_{\text{\ensuremath{\Psi}\ }}=Z_{e}.$ The vacuum polarization diagram
does not diverge in Eq. (\ref{Pi00}), so no divergence is needed
to be canceled by $Z_{\phi}$. Thus, its value is $Z_{\phi}=1.$ The
relation between the bare coupling constant $e_{0}$ and the coupling
constant $e$ is given as
\begin{equation}
e_{0}=Z_{\Psi}^{-1}Z_{e}Z_{\phi}^{-1}e,
\end{equation}
which leads to $e_{0}=e$. It means that the coupling constant $e$
would not be renormalized. Then we use the dimensional regularization
to deal with the divergent one-loop self-energy in Fig. (2a) (see
Appendix \ref{sec:loop})
\begin{equation}
-i\Sigma\left(p\right)=\int J\text{ }\Gamma_{0}G_{e0}\Gamma_{0}G_{\gamma0},
\end{equation}
where the photon's propagator $G_{\gamma0}$ can be read directly
from the Lagrangian in Eq. $\left(\ref{Lagepi}\right)$. Note that
we temporarily ignore the screening effect stemming from finite DOS
in the photon's propagator. There exists a natural momentum cut-off
$\Lambda$ along the $p_{x}$-direction, which characterizes the length
of Fermi lines and is irrelevant to the energy scale. Since the procedure
of renormalization is implemented upon the momentum perpendicular
to Fermi surface rather than the parallel one, $\Lambda$ plays no
role in the renormalization.

By using the $\overline{MS}$ scheme, one can obtain the following
RG equations for the effective velocities and the tilting parameter.
The bare action is given as

\begin{eqnarray}
\mathcal{L} & = & \bar{\Psi}[(Z_{\Psi}p_{0}-Z_{w}w\tilde{w}\eta^{2}v_{y}p_{\perp})\gamma^{0}-\nonumber \\
 &  & Z_{\tilde{w}\eta^{2}v_{y}}\tilde{w}\eta^{2}v_{y}p_{\perp}\gamma^{1}+Z_{v_{y}}v_{y}p_{\perp}\gamma^{2}]\Psi.
\end{eqnarray}
By comparing the bare action with the renormalized one, one gets a
set of equations
\begin{eqnarray}
\Psi_{0} & = & Z_{\Psi}^{1/2}\Psi,\\
v_{y0} & = & Z_{v_{y}}Z_{\Psi}^{-1}v_{y},\\
\left(\tilde{w}\eta^{2}v_{y}\right)_{0} & = & Z_{\tilde{w}\eta^{2}v_{y}}Z_{\Psi}^{-1}\left(\tilde{w}\eta^{2}v_{y}\right),\\
w_{0} & = & Z_{w}Z_{\tilde{w}\eta^{2}v_{y}}^{-1}Z_{\Psi}^{-1}w,
\end{eqnarray}
where $Z_{\Psi}$, $Z_{w}$, $Z_{\tilde{w}\eta^{2}v_{y}}$ and $Z_{v_{y}}$
are given as

\begin{eqnarray}
Z_{\Psi}=Z_{w} & = & 1,\\
Z_{\tilde{w}\eta^{2}v_{y}}=Z_{v_{y}} & = & 1-\frac{e^{2}}{8\pi^{2}\epsilon}F.
\end{eqnarray}
After some standard derivations, one gets the corresponding RG equations

\begin{equation}
\frac{d\ln v_{y}}{d\ln\kappa}=\frac{d\ln(\tilde{w}\eta^{2}v_{y})}{d\ln\kappa}=-\frac{d\ln w}{d\ln\kappa}=\mathcal{B},\label{rgeq3}
\end{equation}
where $\kappa$ is the renormalization scale, i.e., the larger $\kappa$,
the smaller the energy scale. The function $\mathcal{B}\left(w,\ v_{x},\ v_{y}\right)$
is defined by

\begin{eqnarray}
\mathcal{B}\left(w,v_{x},v_{y}\right) & =\frac{e^{2}}{4\pi^{2}} & \frac{\sqrt{1+\tilde{w}^{2}\eta^{2}}}{\sqrt{v_{y}^{2}\left(1+\tilde{w}^{2}\eta^{4}\right)}}.
\end{eqnarray}

From these RG equations in Eq. $\left(\ref{rgeq3}\right)$, one immediately
recognizes that $t_{1}=w\tilde{w}\eta^{2}v_{y}$, $t_{2}=wv_{y}$
and $t_{3}=\tilde{w}\eta^{2}$ are not renormalized, that is, independent
of $\kappa$. $w$ does run with energy scale, which implies that
both the tilting of the cone and anisotropic ratio $v_{y}/v_{x}$
would be renormalized. The non-negative $\beta_{v_{y}}$ and $\beta_{\tilde{w}\eta v_{x}}$
above imply a logarithmic divergence of Fermi velocity, which is similar
to graphene \cite{gonzalez1994npb}. Recently, such a velocity enhancement
has been experimentally observed by Shubnikov\textendash de Haas oscillations
in graphene \cite{elias2011nature,Kotov2012RMP} and by site-selective
nuclear magnetic resonance in the weakly tilted compound $\alpha-\left(\mathrm{BEDT-TTF}\right)_{2}\mathrm{I}_{3}$
\cite{Hirata2016NC}. The negative $\beta_{w}$ means that $w$ is
a decreasing function of length scale then leads to a transition from
type-II Dirac fermions to type-I ones. It is consistent with the recent
result in Ref. \cite{Isobe2016PRL}.

As shown above, a mass term for a photon is induced by the chiral
anomaly, $m_{\text{ph}}^{2}=-\Pi^{00}\left(0,\thinspace\vec{p}\right).$
By setting small external momentum ($p_{\perp}\ll m_{\mathrm{ph}}$)
and using the simplified dressed photon propagator
\begin{equation}
G_{\gamma}=\frac{i}{J\left(p_{\parallel}^{2}+p_{\perp}^{2}\right)+p_{z}^{2}+m_{\mathrm{ph}}^{2}},
\end{equation}
it is straightforward to verify that the self energy diagram is proportional
to $\ln(1+C^{2}/m_{\mathrm{ph}}^{2}),$ where $C$ is the energy cut-off.
The reason $G_{\gamma}$ is referred to as the simplified dressed
photon propagator is that, after integrating over $p_{z}$, $G_{\gamma}$
reduces to be an approximation of the dressed photon propagator. In
the low-energy regime with $C\ll m_{\mathrm{ph}}$, the divergent
part vanishes: $-i\Sigma\sim0$, which means that $v_{x}$, $v_{y}$
and $w$ are barely renormalized. There are two important consequences.
First, the logarithmic divergence of Fermi velocity is substantially
suppressed by the screening effect, which differs from the mechanism
in Ref. \cite{Isobe2016PRL}. Second, 2D type-II anisotropic Dirac
fermions can be stabilized by the screening effect. It is the key
finding of this paper. It should be noted that the one-body operators
hardly receive quantum corrections from the dressed Coulomb interaction,
which coincides with our RPA analysis.

However, the aforementioned $\overline{MS}$ scheme is mass-independent
and does not see the mass thresholds. It is the mass of photon in
this case. This decoupling effect (see Figs. (2d) and (2e)) is implemented
by hand \cite{Appelquist1975prd}: the heavy field (photon) is present
at an energy scale higher than the mass of a photon, but it is integrated
out at the low energy scale, that is, the full theory in Eq. (\ref{eqFeynmanRule})
is valid for the whole energy scale, while the effective field theory
$\mathcal{L}_{\mathrm{\mathrm{eff}}}$ only holds for the low-energy
region

\begin{alignat}{1}
\mathcal{L}_{\mathrm{eff}} & =\bar{\Psi}[\left(a_{1}p_{0}-a_{2}w\tilde{w}\eta^{2}v_{y}p_{\perp}\right)\gamma^{0}-a_{3}\tilde{w}\eta^{2}v_{y}p_{\perp}\gamma^{1}\nonumber \\
 & +a_{4}v_{y}p_{\perp}\gamma^{2}]\Psi+\frac{e^{2}b}{m_{\mathrm{ph}}^{2}}\left(\bar{\Psi}\Psi\right)^{2}+\mathcal{O}\left(m_{\mathrm{ph}}^{-4}\right),
\end{alignat}
where the coefficients $a_{1,2,3,4}$ and $b$ are determined by matching
conditions \cite{pich1998arxiv}: at the energy scale $m_{\mathrm{ph}}$,
these two theories in Eq. (\ref{eqFeynmanRule}) with photon field
and $\mathcal{L}_{\mathrm{eff}}$ without the photon field should
give rise to the same S-matrix elements for light-particle scatterings.
At the tree level, we have $a_{1,2,3,4}=1$, $b=1$. Since the operators
$e^{2}b\left(\Psi^{\dagger}\Psi\right)^{2}/m_{\mathrm{ph}}^{2}+\mathcal{O}\left(m_{\mathrm{ph}}^{-4}\right)$
are irrelevant, the corresponding RG equations for $v_{x},\ v_{y},\ w$
equal to zero in the low-energy region. Although there exists marginal
channels with specific momentum exchange, their effect on the one-body
operator can be absorbed by the chemical potential \cite{Shankar1994RMP}.
Furthermore, one-loop corrections do not change the dimension of operator
$\left(\bar{\Psi}\Psi\right)^{2}$, so this operator remains irrelevant
and the argument above is still valid. Hence, anisotropic type-II
Dirac semimetals are stable against Coulomb interaction. This anisotropy
of the Dirac cone can be accessed by a variety of experimental techniques,
such as electronic transport \cite{Tajima2009STAM}, magnetotransport
\cite{Morinari2009MR} and nuclear magnetic resonance \cite{Hirata2016NC}.

\section{conclusions}

In summary, there exists a phase transition from 2D type-II Dirac
fermions to the type-I ones under the unscreened Coulomb interaction.
Because of the interplay between the nontrivial Fermi surface and
the Coulomb interaction, the logarithmic divergence of Fermi velocity
is suppressed in this case. Additionally, the magnitude of the renormalized
tilting parameter $w$ remains larger than 1, indicating that the
Fermi surface of 2D type-II anisotropic Dirac fermions is stable against
the Coulomb interaction. This anisotropy of tilted Weyl node can be
detected by several experimental techniques. The RG analysis here
can be straightforwardly transplanted to the 3D type-II Weyl semimetals.

Z.M. and J.H. would like to thank Cheng-Feng Cai, Hao-Ran Chang, Kai-Liang
Huang, Jing Wang and Jiabin You for valuable discussions. This work
was supported by the Research Grant Council, University Grants Committee,
Hong Kong under Grant No. 17304414.

\appendix

\section{Self-energy and vacuum polarization\label{sec:loop}}

For simplicity, we first transform the Lagrangian into the following
form

\begin{equation}
\mathcal{L}_{e}=\bar{\Psi}[(p_{0}-w\tilde{w}\eta^{2}v_{y}p_{\bot})\gamma^{0}+rp_{\bot}\gamma^{1}]\Psi,
\end{equation}
with
\[
r=\sqrt{v_{y}^{2}\left(\tilde{w}^{2}\eta^{4}+1\right)}.
\]
The fermion's Green's function and the photon's Green's function are
of the form

\begin{eqnarray}
G_{e0} & = & \frac{i}{p_{0}\gamma^{0}-w\tilde{w}\eta^{2}\gamma^{0}+rp_{\perp}\gamma^{1}}\\
G_{\gamma0} & = & \frac{i}{\left(1+\tilde{w}^{2}\eta^{2}\right)\left(p_{\parallel}^{2}+p_{\perp}^{2}\right)+p_{3}^{2}}.
\end{eqnarray}
And the corresponding one-loop self-energy reads

\begin{align}
-i\Sigma= & J\int\frac{d^{4}q}{(2\pi)^{4}}\left(-ie\gamma^{0}\right)G_{e0}\left(q_{0},q_{\perp}\right)\left(-ie\gamma^{0}\right)\nonumber \\
 & \times G_{\gamma0}\left(q_{\parallel},q_{\perp}+p_{\perp},q_{3}\right).
\end{align}
Making use of the Feynman parametrization

\begin{eqnarray}
\frac{1}{AB} & = & \int_{0}^{1}\frac{dx}{\left[xA+(1-x)B\right]^{2}},
\end{eqnarray}
one rewrites the self-energy as follows

\begin{align}
-i\Sigma & =-e^{2}J\int_{0}^{1}dx\left(x^{2}(1-x)\alpha^{2}K\right)^{-1/2}\nonumber \\
 & \times\int\frac{d^{4}l}{(2\pi)^{4}}\frac{l_{0}\gamma^{0}/\sqrt{1-x}+r\left(l_{1}/\sqrt{K}-x\alpha^{2}p_{\perp}/K\right)\gamma^{1}}{(l_{0}^{2}-l_{1}^{2}-l_{2}^{2}-l_{3}^{2}-\triangle^{2})^{2}},\label{SF2}
\end{align}
where $l_{\mu}$ with $\mu=0,1,2,3$ are given as

\begin{eqnarray}
l_{0}^{2} & = & (1-x)(q_{0}+nq_{\perp})^{2}\nonumber \\
l_{1}^{2} & = & K(q_{\perp}+\frac{x\alpha^{2}}{K}p_{\perp})^{2}\nonumber \\
l_{2}^{2} & = & x\alpha^{2}q_{\parallel}^{2}l_{3}^{2}\nonumber \\
l_{3} & = & xq_{3}^{2}\nonumber \\
\triangle & = & \alpha^{2}p_{\perp}^{2}x(1-K^{-1}\alpha^{2}x),
\end{eqnarray}
with

\begin{eqnarray*}
K & = & (1-x)r^{2}+x\alpha^{2}\\
\alpha & = & \sqrt{1+\tilde{w}^{2}\eta^{2}}\\
n & = & -w\tilde{w}\eta^{2}v_{y}.
\end{eqnarray*}
 Since the denominator in the integrand in Eq. $\left(\ref{SF2}\right)$
is an even function of $l_{\mu}$ with $\mu=0,1,2,3$, the terms odd
in $l_{\mu}$ in the numerator make no contribution to the integral.
We then make a Wick's rotation and perform integration over $l_{\mu}$
by using the dimensional regularization

\begin{eqnarray}
\int\frac{d^{d}k}{(2\pi)^{d}}\frac{1}{(k^{2}+\triangle)^{m}} & = & \frac{\Gamma(m-d/2)}{(4\pi)^{d/2}\triangle^{m-d/2}\Gamma(m)},
\end{eqnarray}
finally reach the self-energy

\begin{align}
-i\Sigma & =\frac{ie^{2}}{16\pi^{2}}F\left(w,v_{x},v_{y}\right)\left(rp_{\perp}\gamma^{1}\right)\nonumber \\
 & \times\Gamma\left(\frac{\epsilon}{2}\right)\left(\frac{1}{\triangle}\right)^{\epsilon/2},
\end{align}
where the function $F\left(w,v_{x},v_{y}\right)$ is defined as
\begin{equation}
F\left(w,v_{x},v_{y}\right)=2\sqrt{1+\tilde{w}^{2}\eta^{2}}/\sqrt{v_{y}^{2}\left(\tilde{w}^{2}\eta^{4}+1\right)}.
\end{equation}
Carrying out a inverse transformation leads us to find
\begin{align}
-i\Sigma & =\frac{ie^{2}}{16\pi^{2}}F\left(w,v_{x},v_{y}\right)\left(-\tilde{w}\eta^{2}\gamma^{1}+\gamma^{2}\right)\nonumber \\
 & \times v_{y}p_{\perp}\Gamma\left(\frac{\epsilon}{2}\right)\left(\frac{1}{\triangle}\right)^{\epsilon/2}.
\end{align}
In the limit of $\epsilon\rightarrow0$, one finds
\[
\Gamma\text{(}\frac{\epsilon}{2}\text{)}=\frac{2}{\epsilon}-\gamma+\mathcal{O}(\epsilon^{2}),
\]
and
\[
(\frac{1}{\triangle})^{\epsilon/2}=1-\frac{\epsilon}{2}\ln\triangle+\mathcal{O}(\epsilon^{2}).
\]
Thus the divergent part of self-energy $-i\Sigma$ becomes

\begin{eqnarray}
-i\Sigma & \rightarrow & \frac{ie^{2}p_{\perp}}{8\pi^{2}\epsilon}F\left(w,v_{x},v_{y}\right)v_{y}\left(-\tilde{w}\eta^{2}\gamma^{1}+\gamma^{2}\right).
\end{eqnarray}
The vacuum polarization diagram can be evaluated in a similar manner

\begin{align}
i\Pi^{00} & =-J\int\frac{d^{4}q}{(2\pi)^{2}}\text{tr}\left[(-ie\gamma^{0})G_{e}(q_{0},q_{\perp})\right.\nonumber \\
 & \times\left.(-ie\gamma^{0})G_{e}(q_{0}+p_{0},q_{\perp}+p_{\perp})\right]\nonumber \\
\nonumber \\
 & =\frac{-iJe^{2}}{\pi\sqrt{r^{2}}}\left[\frac{(p_{0}+np_{\perp})^{2}+r^{2}p_{\perp}^{2}}{(p_{0}+np_{\perp})^{2}-r^{2}p_{\perp}^{2}}-1\right]\Lambda^{2},
\end{align}
where $\Lambda$ is the cut-off along the $q_{3}$-direction and
\[
\mathcal{A}=\int\frac{d^{2}l}{(2\pi)^{2}}\frac{l_{0}^{2}+l_{1}^{2}}{\left(l_{0}^{2}-l_{1}^{2}-\triangle\right){}^{2}}=-\frac{i}{4\pi}.
\]
 Note that we have used Feynman parametrization in the second step
and carried out Wick's rotation in the last step.

\end{document}